\documentclass{pasa}%

\usepackage{graphicx}

\title[CH with an SKA Precursor]{First Search for Low-Frequency CH with a Square Kilometre Array Precursor Telescope}

\author[Tremblay et al. ]{Tremblay, C.D.$^1$, Green, J.A. $^1$, Mader, S.L.$^{2}$, Phillips, C.J.$^{3}$ and Whiting, M.$^{3}$
\affil{$^1$CSIRO Astronomy and Space Science, PO Box 1130, Bentley WA 6102, Australia}%
\affil{$^2$CSIRO Astronomy and Space Science, Parkes Radio Telescope, NSW, Australia}
\affil{$^3$CSIRO Astronomy and Space Science, PO Box 76, Epping NSW 1710, Australia}
}%

\jid{PASA}
\doi{10.1017/pas.\the\year.xxx}
\jyear{\the\year}

\usepackage{aas_macros}
\usepackage{hyperref} 
\hypersetup{colorlinks,citecolor=blue,linkcolor=blue,urlcolor=blue}

\hypersetup{draft}

\begin{document}

\begin{frontmatter}
\maketitle

\begin{abstract}
The diatomic free radical methylidyne (CH) is an important tracer of the interstellar medium and the study of it was critical to our earliest understanding of star formation.  Although it is detectable across the electromagnetic spectrum, observations at radio frequencies allow for a study of the kinematics of the diffuse and dense gas in regions of new star formation. There is only two published (single-dish) detections of the low-frequency hyperfine transitions between 700 and 725\,MHz, despite the precise frequencies being known. These low-frequency transitions are of particular interest as they are shown in laboratory experiments to be more sensitive to magnetic fields than their high-frequency counterparts (with more pronounced Zeeman splitting).  In this work we take advantage of the radio quiet environment and increased resolution of the Australian Square Kilometre Array Pathfinder (ASKAP) over previous searches to make a pilot interferometric search for CH at 724.7883\,MHz (the strongest of the hyperfine transitions) in RCW 38.  We found the band is clean of radio frequency interference, but we did not detect the signal from this transition to a five sigma sensitivity limit of 0.09\,Jy, which corresponds to a total column density upper limit of 1.9$\times$10$^{18}$\,cm$^{-2}$ for emission and 1.3$\times$10$^{14}$\,cm$^{-2}$ for absorption with an optical depth limit of 0.95. Achieved within 5\,hrs of integration, this column density sensitivity should have been adequate to detect the emission or absorption in RCW 38, if it had similar properties to the only previous reported detections in W51. 

\end{abstract}


\end{frontmatter}

\section{INTRODUCTION }
\label{sec:intro}

\label{S:1}
Free radicals are molecules that are uncharged, highly reactive and short lived.  In the early 1920s emission bands due to CH, OH, and CN radicals were first discovered in the laboratory, then in 1937 the diatomic free radical of methylidyne, CH, was discovered \citep{Swings_1937,Swings_1942} as one of the first gas phase molecules. In 1941, a group of Belgian astronomers worked with laboratory chemists to identify CH in the optical spectra of comets (Cunningham 1940c).  This sparked an interest by laboratories and astronomers alike in CH \citep{Herzberg_Spectra} and  since the 1940's, scientists have suggested that there are two forms of CH found around interstellar objects; the molecules that form on grains and are observed in ultra-violet and infrared through the vibrational bands (although never detected); and the molecules that form in the gas phase and are observed in the ultra-violet to the radio parts of the electromagnetic spectrum.

The production of CH is expected to form in the gas phase through a radiative association reaction of C$^{+}$ + H $\rightarrow$ CH + $h\nu$ \citep{Federman_Diffuse}. \cite{Danks_CH} suggested a more thorough description of the formation of diffuse interstellar CH by having H$_{2}$ playing a major role: by following the same radiative association reaction with a ion-molecule reaction CH$_{2}^{+}$ + H$_{2}$ $\rightarrow$ CH$_{3}^{+}$ + H with a 67\,per\,cent efficiency at creating a CH molecule. It is now well accepted that CH originates from the cold neutral medium of interstellar gas via the slow radiative association of C$^{+}$ with H$_{2}$ forming CH$^{+}_{2}$, which converts to CH via dissociative recombination \citep{Godard_2014,Gerin-2016}. The rate constants for the radio frequencies are not well understood \citep{Dagdigian_2018}. The chemical models of this reaction can produce the column densities of CH as observed in the gas phase \citep{Crawford_CH, Gerin-2016} and found to be highly dependant on the presence of free carbon \citep{Suutarinen_2011}. In work published by \cite{Dagdigian_2018}, they found that when investigating collision rates with atomic and molecular hydrogen the column densities from historical observations do not fit the model if local thermodynamic equilibrium (LTE) conditions are assumed, except in regions where the gas is dense. They also found that the calculations using only excitation by H$_2$ do not reproduce the excitation pattern of the lamda doubling transitions near 3.3\,GHz. 


There have been a number of publications regarding CH; its formation routes \citep[e.g.][references therein]{Millar_2015}, its role as a proxy for H$_{2}$ (e.g. \citealt{Sheffer_2008,Wiesemeyer_CH,Dailey_2020}), and its ability to trace the dynamics of diffuse and dense gas (e.g. \citealt{Federman_Diffuse, Gerin-2016}).  Since CH is found in H{\sc i} regions, H{\sc ii} regions, and dust clouds, there is reason to believe the molecule is widespread within our Galaxy \citep{Rydbeck_Radio}  and has been widely observed in diffuse interstellar gas clouds at 3\,GHz \citep{Chastain_2010}. 

A recent resurgence of the study of CH has started with laboratory work by \cite{Truppe_CHMolecule}, completing a focused study on the 700 and 3300\,MHz radio transitions and \cite{Wiesemeyer_CH} completing a study of the N$=$2 ground state line in the infrared.  \cite{Truppe_CHMolecule} combined laboratory experiments with past astronomical observations to determine the precise frequencies of emission, the interaction with the Zeeman effect  and the use of CH to determine fundamental universal constants. 
The interaction with the Zeeman Effect (the splitting of transitions due to the presence of a magnetic field) is of particular interest as it has been found that these transitions have a splitting factor comparable to that of mainline OH, 0.93\,Hz\,$\mu$G$^{-1}$, and thus has the potential to allow for this molecule to trace the impact of magnetic fields on the chemical evolution of the interstellar medium (see \citealt{Crutcher_2019} for a full review on the importance of Zeeman Effect in star formation).
\cite{Wiesemeyer_CH} used observations at 2\,THz to compare CH with traditional proxies of molecular hydrogen in cold gas and determined that CH provided more information over traditional models, including CO.

Based on its expected abundance and appearance in diffuse and dense gas, analysis of CH at low radio frequencies is an ideal target for the Square Kilometre Array (SKA) to study the role magnetic fields play in the early stages of star formation.  The Australian Square Kilometre Array Pathfinder (ASKAP), with operational frequencies of 700--1800\,MHz and stationed in a protected radio quiet environment, makes for an ideal precursor instrument to determine the feasibility of these low frequency studies. This paper presents the first interferometric search using early science observations with ASKAP.   

\section{Previous Radio Observations of CH}

Most of the detections and study of the CH radical are in the ultraviolet (e.g. \citealt{Herzberg_Spectra}), optical (e.g. \citealt{Danks_CH}) and infrared (e.g. \citealt{Wiesemeyer_CH}) parts of the electromagnetic spectrum.  The first radio detection of interstellar CH (in the 3.3 GHz transitions, once observational wavelengths were known) was toward seventeen H{\sc i} and H{\sc ii} regions (almost all with previous OH detections) by \cite{Rydbeck_Radio} and \cite{Federman_CH}. \cite{Rydbeck_Radio} found all detections were in emission and toward OH maser sites, with the CH transition showing evidence of weak masing toward some sources suggesting CH can be detected where the lower rotational lines of CO could not. 

A search for CH (3.3GHz) along the Galactic plane by Johansson (1979) found the molecular distribution more extended than CO(1-0) surveys of the time.
Additional studies of the CH 3\,GHz distribution towards small dark molecular \citep{Federman_CH, Mattila_CH} and high latitude translucent \citep{Magnani_CH1, Chastain_2010, Magnani_CH2} clouds have shown the CH molecule to be an excellent tracer for H$_{2}$ in low-density (n $\leq$ 10$^{3}$ cm$^{3}$; A$_{V}$ $\leq$ 5 mag.) gas, with a linear relationship existing between N(CH) and N(H$_{2}$) for these regions.  \cite{Dailey_2020} studied the possibility of using the 3\,GHz transitions as a proxy of the molecular hydrogen gas and found that the excitation temperatures vary significantly depending on the line-of-sight.

Whilst the detection of the 3335 MHz transition of CH is readily obtained across a wide range of molecular environments, there are very few published detection of CH at the lower transitions of $\sim$ 700 MHz.
\cite{Ziurys_CH} observed the F$=$2--2 and F$=$1--1, N$=$1, J$=$3/2 transitions between 700 and 725\,MHz with the Arecibo 300m telescope and NRAO Green Bank 300\,foot (91m) telescope followed by \cite{Turner_1988}.  \cite{Ziurys_CH} detected two lines in absorption toward W51 and the F$=$2--2, N$=$1, J$=$3/2 transition was reported as detected toward W3, W43, and Ori B in absorption with Arecibo.  The lines detected toward W51A showed a matching profile to the 3.3\,GHz observations by \cite{Genzel_CH_Radio}. \cite{Turner_1988} detected all four hyperfine transitions toward W51A. \cite{Ziurys_CH} found large fractional abundances of CH compared to expectations from chemical modelling of the W51 region. Given the lack of observations at the lower frequencies, more data is required to ascertain the extent of the W51 results across other regions.

\section{New Radio Observations of CH}
\begin{figure}
	\centering 
	 \includegraphics[width=0.48\textwidth]{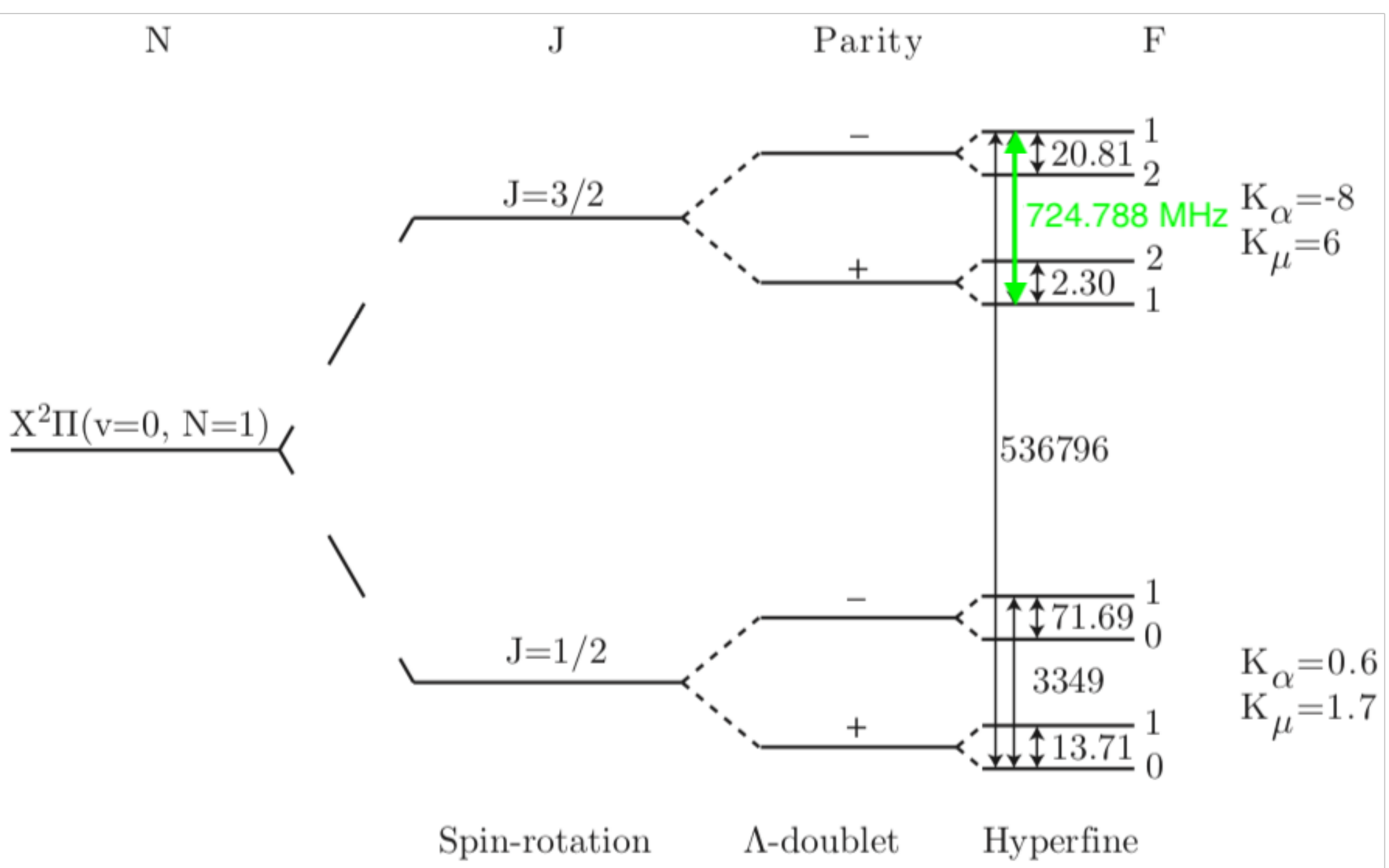}
     \caption{Relevant energy levels in CH from Figure 1 of \cite{Truppe_CHMolecule}. Approximate frequencies of separation are given in MHz, with the 724.788315\,MHz transition highlighted in green. The sensitivity coefficients (K) for the two $\lambda$-doublet transitions are shown.}
\label{energy}
\end{figure}

The CH molecule has four hyperfine transitions in the range $\sim$700--725\,MHz representing the N=1 J=3/2 transition.  The 724.788\,MHz transition (\textbf{F$=$1--1}) is shown in Figure\,\ref{energy} along with the higher frequency N=1 J=1/2 transition at 3349\,MHz.

The transition at 701.6777\,MHz was found to be the strongest by \cite{Ziurys_CH} followed by the transition at 724.7883\,MHz.  The intervening transitions at 722.4876 and 703.9783\,MHz were weakly detected by \cite{Turner_1988}. These low-frequency hyperfine transitions have been found to be both weaker (1/25th the sensitivity) and less prevalent (5\% of molecules) than the more well studied $\sim$ 3.3\,GHz J=1/2 transitions \citep{Truppe_CHMolecule}. Observations with the Effelsberg 100\,m telescope of the J=1/2 transitions by \cite{Genzel_CH_Radio} detected the lines in H{\sc ii} regions with 2--3 hours of integrated time, thus implying a significant integration time is required for the J=3/2 transitions (consistent with the observations by \citealt{Ziurys_CH} where they specify observing across several days). New studies of these transitions therefore need to exploit the most sensitive new telescopes, such as the Australian Square Kilometre Array Pathfinder.

\subsection{Australian Square Kilometre Array Pathfinder (ASKAP)}
\begin{table}
\small
\caption{ASKAP Observing Parameters}
\label{obs}
\begin{tabular}{ll}
\hline
Parameter & Value\\
\hline
\hline
Central frequency& 724\,MHz\\
Total bandwidth & 18\,MHz\\
Channel Resolution & 1\,kHz \\
Synthesised beam FWHM & 30$^{\prime\prime}$\\
Primary beam FWHM$^{\star}$ & 30\, deg$^{2}$\\
Phase centre of image (J2000) & 09h59m05s \\
&--47d30m39s\\
Time on source & 5\,hours \\
Reference Frame & Local Standard\\
\hline

\end{tabular}\\
$\star$The primary beam size represents the size of a full processed field with 36 PAF beams. A single PAF beam has a primary beam of 1\,deg$^{2}$.
\end{table}

The Australian Square Kilometre Array Pathfinder (ASKAP; \citealt{ASKAP_Science}, Hotan et al. submitted) is located at the Murchison Radio-astronomy Observatory (MRO) in Western Australia.  Observations were taken on the 23 February 2019 with a central frequency of 724\,MHz and 18\,MHz of total bandwidth, with a frequency resolution of 1\,kHz. This provides a 0.25\,km\,s$^{-1}$ velocity resolution.  Each of the antennas are equipped with a Phased Array Feed (PAF; \citealt{Schinckel-PAF}) giving it a field of view, over 36 beams, of $\sim$ 30 deg$^{2}$. 

The observations were taken as part of the commissioning of the low-band of ASKAP \citep{McConnell_2016} and 35\footnote{Antenna 11 was offline for these observations.} antennas were online with a minimum baseline of 22\,m and maximum baseline of 6\,km. The observations were taken in full polarisation and observations of PKS B1934--638 were performed immediately adjacent in time to the target field for the purposes of bandpass calibration. Each calibration observation contained one calibrator scan of duration five minutes at the centre of each PAF beam before mosiacing to the full instrument field-of-view (FOV).

All data were processed using ASKAP{\sc soft} \citep{Guzman_Askapsoft} for calibration and imaging.  This included flagging, bandpass calibration, phase self-calibration, imaging, continuum subtraction, spectral cube imaging and correction of the data to the local standard of rest reference frame.  During processing antennas 1, 4 and 10, from the inner core, were flagged\footnote{ Information on the antenna layout and configuration is found at https://www.atnf.csiro.au/projects/askap/config.html.}.  This combined with antenna 11 being offline, reduced our surface brightness sensitivity by $\sim$40\,per\,cent and thus our sensitivity to diffuse emission. Nine PAF beams were processed individually and then mosaiced together to form an image of the region used in this paper. However, the H{\sc ii} region of interest fit within a single central beam. 

\subsection{RCW 38: H{\sc ii} Region}

We completed a targeted survey with ASKAP toward RCW 38 to look for the low-frequency transition of CH at 724.788\,MHz with five hours of integration time and a pointing centre of RA 09:59:05 Dec --47:30:39 (J2000).  The molecule CH has been detected in a number of different astrophysical environments but the only detection of a low-frequency transition published was toward the OH maser site of W51A \citep{Ziurys_CH}. As RCW 38 is also an H{\sc ii} region thought to be fuelled by an old supernova explosion and colliding clouds (e.g.\citealt{Wolk_2006,Schneider2010,Sano2018}), we completed our first attempt to search for the molecule with ASKAP toward this region.

\begin{figure*}
\centering
\includegraphics[width=0.39\textwidth]{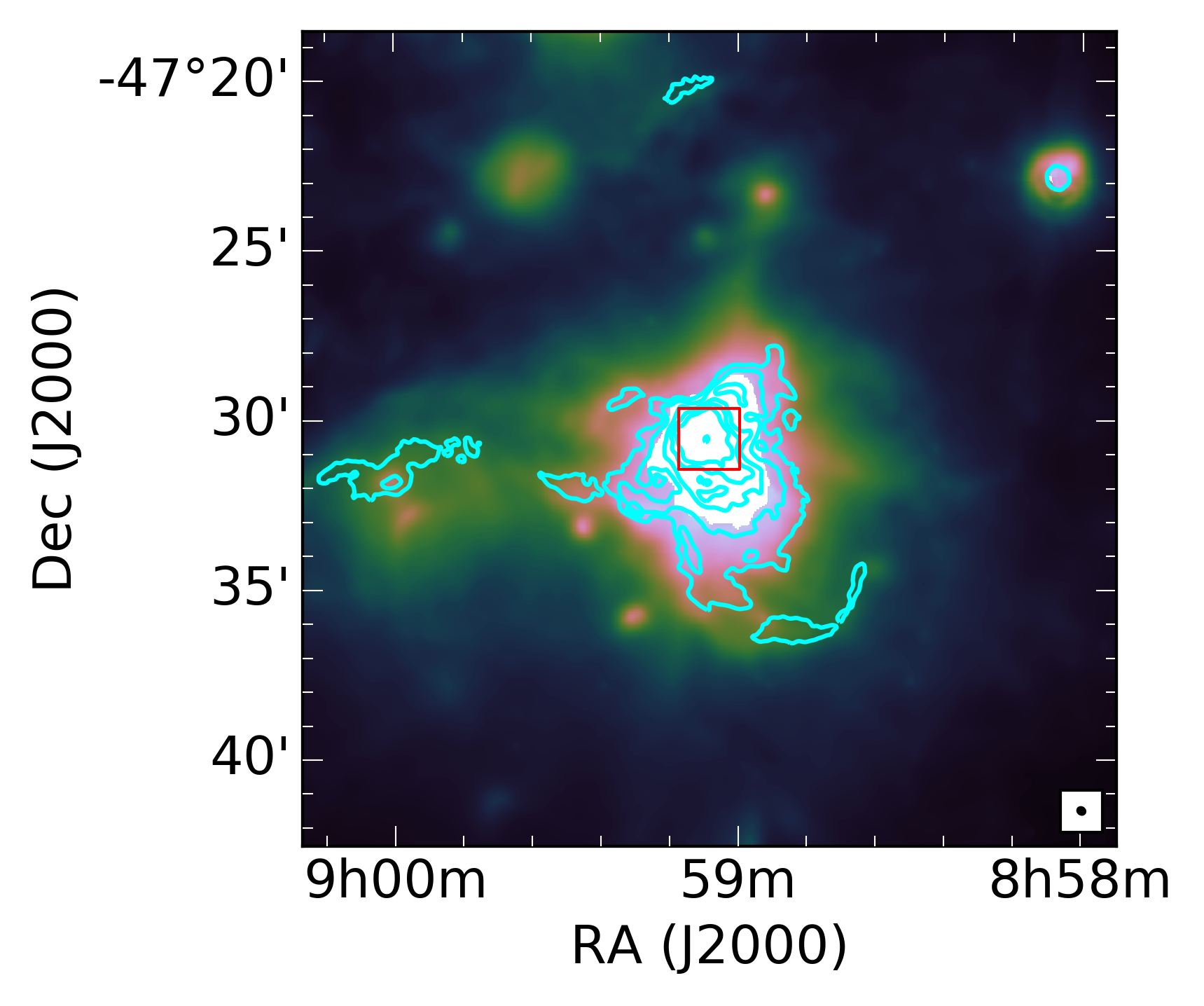}
\includegraphics[width=0.60\textwidth]{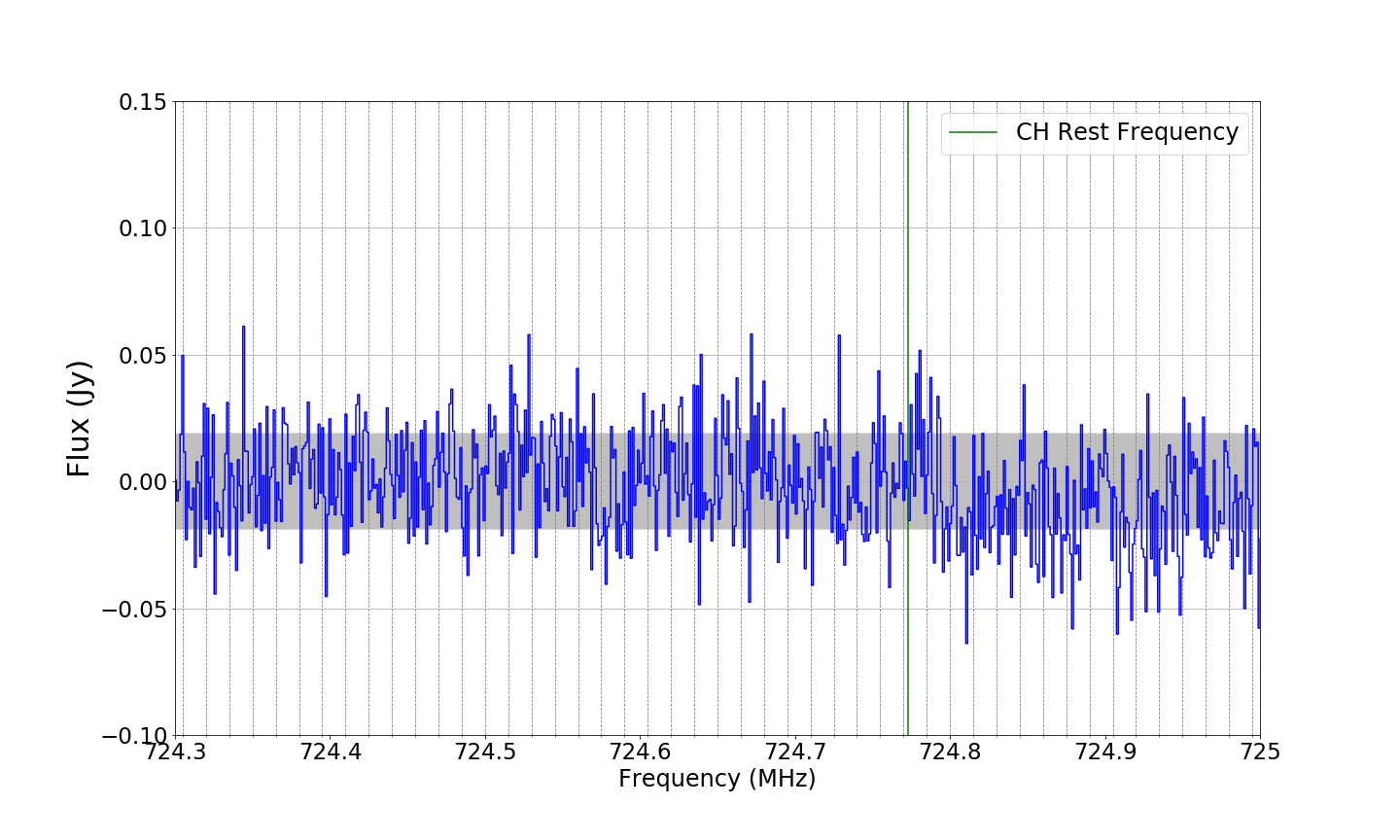}
\caption{Left: An image of the eastern lobe (IRS1) of RCW 38, where the cyan contours from the ASKAP radio continuum emission are overlaid onto an infrared image of the region from the Wide-field Infrared Survey Explorer (WISE) survey. The contours show greater detail of the lobes typically denoted as ``IRS1'' in the literature \citep{WynnWilliams1972}. The contours are set at levels of 3--8\,$\sigma$ (0.003 to 0.09\,Jy\,beam$^{-1}$) and the bottom right-hand corner shows the size of the synthesised beam. Right: The spectrum averaged over the region marked with a red box on the image on the left. The rest frequency of the CH transition (724.7883 MHz) is denoted by a green vertical line, the grey shading represents noise (one sigma) and the five sigma sensitivity limit is taken as 0.09\,Jy.}
\label{RCW39radio}
\end{figure*}

RCW 38, an H{\sc ii} region located in the Vela Molecular Cloud Complex, is 1.7$\pm$0.9\,kpc away \citep{Wolk_2006}, and is nearly as dense as the Orion Kleinmann-Low Nebula with young stellar objects \citep{Schneider2010}.  See \cite{Wolk_2008} for a review of the region and the historical observations. \cite{Torii_2019}recently completed a high-resolution survey of CO and its isotopologues with the Atacama Large Millimeter/submillimeter Array. They detected CO in the lobes of RCW 38, often referred to as IRS1 and IRS2 \citep[from][]{WynnWilliams1972}, and the bridge of gas that expands between the two with two different radial velocities of $\sim$ 2 and $\sim$ 12\,km\,s$^{-1}$.  \cite{Bourke_2001} detected the 1665 and 1667 OH lines in absorption with detectable Zeeman Splitting (with a corresponding magnetic field strength of 38$\pm$3 $\mu$G).  The source is often characterised as being two clouds colliding which fuels the significant star formation with over 1000 objects detected by X-ray and infrared (e.g. \citealt{Fukui2016,Sano2018}).  In Figure \ref{RCW39radio}, the ASKAP radio continuum is shown in contours over an infrared image from the Wide-field Infrared Survey Explorer (WISE) Survey.  The contours reveal new structure to the IRS1 cloud but is found to be similar to the structure detected by \cite{Bourke_2004} at 1.6\,GHz with the Australia Telescope Compact Array.  Greater detail regarding this new structure will be presented in a future paper.

\section{Results and limits for detection}

As shown in the spectrum on the right-hand side of Figure \ref{RCW39radio}, we did not detect the CH transition towards RCW 38 with our ASKAP observations to a five sigma limit of 0.09\,Jy after averaging the signal over the region marked by the red box in the left-hand image of the eastern lobe of RCW38.  

We calculated the upper limit in emission (at five sigma) of the total column from Equation A1 of \cite{Tremblay_2017} at a T$_\mathrm{{ex}}$ of 50\,K to be 1.9$\times$10$^{18}$\,cm$^{-2}$. \cite{Ziurys_CH} determined the column density derived for the lower energy state to be 9.8$\times$10$^{17}$\,cm$^{-2}$ for CH in W51A, assuming similar physical conditions. However, \cite{Turner_1988} suggests this calculation may be simplistic and not a good limit on the gas content.

The previous detections of the low-frequency CH lines by \cite{Ziurys_CH} and \cite{Turner_1988} toward W51 were both in absorption. Using the spectrum in Figure \ref{RCW39radio}, we can make some estimate of the optical depth in relation to the peak intensity in the continuum. For this we get an optical depth limit of 0.95.  If we use the modified column density Equations 1 and 2 from \cite{Turner_1988} we get a five-sigma upper limit on the column density of 1.3$\times$10$^{14}$\,cm$^{-2}$ which is consistent with historical values in W51 and other H{\sc ii} regions (around 1$\times$10$^{14}$\,cm$^{-2}$ \citealt{Genzel_CH_Radio,Rydbeck_Radio}).  However, as noted by both \cite{Dailey_2020} and \cite{Dagdigian_2018}, the values of column density are likely non-physical measurements as the radio lines do not follow LTE conditions.


We do note that the previously reported low-frequency detections of CH was achieved using the 300\,m Arecibo telescope, which as a single dish with a beam size of more than 6.3\,arc\,minutes and has greater sensitivity to diffuse (low surface brightness) emission. In comparison, a single ASKAP's PAF equipped on the antennas (at these frequencies) have a primary beam of $\sim$1\,degree, and the resultant imaging has a 30\,arc\,second synthesised beam. In these observations we lost a large percentage of the shorter baselines, thus making us less sensitive to the diffuse gas but this will not effect sensitivity to the dense gas. \cite{Danks_CH} noted that the signal from the diffuse gas was stronger than that from the dense gas and the two components are easily differentiated through resolved velocity components. 

As such, observations with the full ASKAP array of 36 antennas, with all of the shorter baselines, may be required for future success. Additionally, we note that \cite{Ziurys_CH} quote a velocity resolution 1.4\,km\,s$^{-1}$ whereas our observations have a velocity resolution of 0.25\,km\,s$^{-1}$.  With the Parkes 64m Telescope fitted with the new Ultra-wideband receiver \citep{Hobbs_UWL}, we detected CH in the eastern lobe (IRS1) at 3.264\,GHz in a 10\,minute snapshot image.  Therefore, we know the molecule to exist in the region and more observing time should yield a detection of the low-frequency transitions.  

\section{Implications for future studies}
Based on the current presented observations, the best chance for detection of the J=3/2 transition at 724.788\,MHz is to obtain observations for at least 8 hours of integration with ASKAP (using all 36 antennas). However, this is the first published search for CH with an interferometer. In the previous single-dish searches the beam filling factor is assumed to be 1, so interferometry will help determine if this is true \citep{Wiesemeyer_CH}. It will also help identify if the excitation temperature is indeed negative.  This may impact the length of observation required to detect CH with an interferometer such as ASKAP.  

It is also important to further consider the choice of target: although we now have a wealth of information on the astrophysical properties of regions from Galactic plane surveys, which can form the basis of targeted observations, all of the previously published CH detections are towards OH maser sites. The combination of OH and CH (with their respective broad Zeeman splitting factors) can provide a useful diagnostic of magnetic fields in differing stages of the interstellar medium.  Surveys of OH that are planned with ASKAP \citep{GASKAP} along the Galactic plane and long-term surveys like the The Southern Parkes Large-Area Survey in Hydroxyl (SPLASH; \citealt{Dawson_SPLASH}) may provide the best targets to be used for understanding CH and its role in the interstellar medium.

Both of the components (mid- and low-frequency arrays) of the future Square Kilometre Array (SKA) telescope will be in the southern hemisphere, whereas most of the searches for CH at radio wavelengths have been towards targets accessible by the northern hemisphere. As such the precursors like ASKAP can provide us with useful regions to target later with the SKA.

The CSIRO Parkes 64m Telescope, an SKA technology pathfinder, has a new ultra-wide band receiver \citep{Hobbs_UWL}, which covers the frequency range of 704--4000\,MHz, allowing simultaneous studies of the 724.788\,MHz and $\sim$3.3\,GHz CH transitions.  \cite{Rydbeck_Radio} observed CH ground state transitions at 3.264\,GHz in seventeen HII regions. They found a resemblance of CH to the OH excited state of 2$\Pi$1/2, J=1/2 transitions, which can also be observed with Parkes.  

Searches for other low-frequency molecular transitions associated with CH may also prove to be beneficial to understanding the interstellar medium.  \cite{Rydbeck_Radio} found that CH velocities matched those of formaldehyde, an important tracer of core collapse in early high-mass stars. Both formaldehyde and deuterated formaldehyde have transitions detectable by the Murchison Widefield Array (MWA; also a SKA precursor) between 70--300\,MHz and deuterated formaldehyde was found in the Orion complex with the MWA by \cite{Tremblay_2018}.

A further consideration for future studies, as alluded to earlier, is that \cite{Truppe_CHMolecule} found that the 722--724\,MHz transitions of CH to be particularly sensitive to magnetic fields and the line shows polarisation, which ASKAP is capable of detecting.  As CH is detectable in diffuse and dense gas, this provides a new way of studying the role magnetic fields play in the early stages of star formation in comparison to OH which is only detectable in dense gas.

\section{Conclusion}
Using the Australian Square Kilometre Array Pathfinder telescope, an SKA precursor in Western Australia, we have completed the first interferometric search of the low frequency transition of the CH molecule at 724.7883 MHz towards RCW 38. We established a five sigma upper limit for non-detection of 0.09\,Jy, which corresponds to a total column density upper limit of 1.9$\times$10$^{18}$\,cm$^{-2}$ for emission and 1.3$\times$10$^{14}$\,cm$^{-2}$ for absorption with an optical depth limit of 0.95. Based on the column density limit of previous detection of CH, this should have been adequate for detection.   These limits, combined with a spectrum free from radio frequency interference, show that detection of the 724\,MHz transition should be possible with ASKAP with an integration time of $\sim$8\,hours, but would benefit from having the full 36 antenna array to increase the sensitivity to the diffuse gas. 

\begin{acknowledgements}
We thank the anonymous referee for their comments and assistance on improving this manuscript. This scientific work makes use of the Murchison Radio-astronomy Observatory, operated by CSIRO. We acknowledge the Wajarri Yamaji people as the traditional owners of the Observatory site. Operation of ASKAP is funded by the Australian Government with support from the National Collaborative Research Infrastructure Strategy. Establishment of ASKAP, the Murchison Radio-astronomy Observatory and the Pawsey Supercomputing Centre are initiatives of the Australian Government, with support from the Government of Western Australia and the Science and Industry Endowment Fund.  This research has made use of NASA’s Astrophysics Data System Bibliographic Services. We acknowledge the Pawsey Supercomputing Centre which is supported by the Western Australian and Australian Governments. Access to Pawsey Data Storage Services is governed by a Data Storage and Management Policy (DSMP).  This publication makes use of data products from the Wide-field Infrared Survey Explorer, which is a joint project of the University of California, Los Angeles, and the Jet Propulsion Laboratory/California Institute of Technology, funded by the National Aeronautics and Space Administration.
\end{acknowledgements}

\bibliographystyle{pasa-mnras}
\bibliography{CH}

\end{document}